\documentclass[conference]{IEEEtran}
\usepackage{graphicx}

\begin{document}

\title{Level Crossing Rate and Average Fade Duration \\ of the Multihop Rayleigh Fading Channel}
\author{\IEEEauthorblockN{Zoran Hadzi-Velkov}
\IEEEauthorblockA{Faculty of Electrical Engineering and \\Information Technologies, Ss. Cyril \\and Methodius University, Skopje\\
Email: zoranhv@feit.ukim.edu.mk \vspace{-0.5mm}} \and
\IEEEauthorblockN{Nikola Zlatanov}
\IEEEauthorblockA{Faculty of Electrical Engineering and \\Information Technologies, Ss. Cyril \\and Methodius University, Skopje\\
Email: nzlatanov@manu.edu.mk \vspace{-0.5mm}} \and
\IEEEauthorblockN{George K. Karagiannidis}
\IEEEauthorblockA{Department of Electrical and Computer \\Engineering, Aristotle University of \\Thessaloniki, Thessaloniki\\
Email: geokarag@auth.gr \vspace{-0.5mm}}} \maketitle

\vspace{-1.5cm}
\begin{abstract}
We present a novel analytical framework for the evaluation of
important second order statistical parameters, as the level
crossing rate (LCR) and the average fade duration (AFD) of the
amplify-and-forward multihop Rayleigh fading channel. More
specif\-ically, motivated by the fact that this channel is a
cascaded one, which can be modelled as the product of $N$ fading
amplitudes, we derive novel analytical expressions for the average
LCR and AFD of the product of $N$ Rayleigh fading envelopes, or of
the recently so-called $N*$Rayleigh channel. Furthermore, we
derive simple and eff\-icient closed-form approximations to the
aforementioned parameters, using the multivariate Laplace
approximation theorem. It is shown that our general results reduce
to the specif\-ic dual-hop case, previously published. Numerical
and computer simulation examples verify the accuracy of the
presented mathematical analysis and show the tightness of the
proposed approximations.
\end{abstract}

\vspace{-0.5cm} \footnote{Accepted at IEEE ICC 2008}
\section{Introduction}
\vspace{-0.3cm} \PARstart{M}{ultihop} communications, a viable
option for providing broader and more eff\-icient coverage, can be
categorized as either non-regenerative (amplify-and-forward, AF)
or regenerative decode-and-forward, DF) depending on the relay
functionality \cite{1}-\cite{Patel}. In DF systems, each relay
decodes its received signal and then re-transmits this decoded
version. In AF systems, the relays just amplify and re-transmit
their received signal. Furthermore, a system with AF relays can
use channel state information (CSI)-assisted relays \cite{1} or
f\-ixed-gain relays \cite{2} (also known as blind or semi-blind
relays \cite{9}). A (CSI)-assisted relay uses instantaneous CSI of
the channel between the transmitting terminal and the receiving
relay to adjust its gain, whereas a f\-ixed-gain relay just
amplif\-ies its received signal by a f\-ixed gain
\cite{2}\cite{9}. Systems with f\-ixed-gain relays perform close
to systems with (CSI)-assisted relays \cite{2}, while their easy
deployment and low complexity make them attractive from a
practical point of view.

Several works in the open literature have provided performance
analysis of AF or DF systems in terms of bit error rate (BER) and
outage probability under different assumptions of the amplif\-ier
gain \cite{1}-\cite{Patel}. Among them, only two works dealt with
the dynamic, time-varying nature of the underlying fading channel,
\cite{Yang}, \cite{Patel}, despite the fact that it is necessary
for the system's design or rigorous testing. In \cite{Yang}, the
level crossing rate (LCR) and the average fade duration (AFD) of
multihop DF communication systems over generalized fading channels
was studied, both for noise-limited and interference-limited
systems, while Patel et. al in \cite{Patel} provide useful exact
analytical expressions for the AF channel's temporal statistical
parameters such as the auto-correlation and the LCR. However, the
approach presented in \cite{Patel} is limited only to the dual-hop
f\-ixed-gain AF Rayleigh fading channel.

In this paper, we study the second order statistics of the
f\-ixed-gain AF multihop Rayleigh fading channel. More
specif\-ically, motivated by the fact that this channel is a
cascaded one, which can be modeled as the product of $N$ fading
amplitudes, we derive a novel analytical framework for the
evaluation of the average LCR and the AFD of the product of $N$
Rayleigh fading envelopes. Furthermore, we derive simple and
eff\-icient closed-form approximations using the multivariate
Laplace approximation theorem [16, Chapter IX.5], \cite{18}. These
important theoretical results are then applied to investigate the
second order statistics of the multihop Rayleigh fading channel.
Numerical and computer simulation examples verify the accuracy of
presented mathematical analysis and show the tightness of the
proposed approximations.


\section{Level Crossing Rate and Average Fade Duration of The Product of $N$ Rayleigh Envelopes }

Let $\{{X_i(t)}\}_{i=1}^N$ be $N$ independent and not necessarily
identically distributed (i.n.i.d.) Rayleigh random processes, each
distributed according to \cite{11}-\cite{12},
\begin{equation}\label{1}
f_{X_i}(x)=\frac{2x}{\Omega_i} \exp \left(
-\frac{x^2}{\Omega_i}\right) , \qquad x \geq 0,
\end{equation}
in an arbitrary moment $t$, where $\Omega_i=E\{X_i^2(t)\}$ is the
mean power of the $i$-th random process ($1\leq i \leq N$).

If $\{{X_i(t)}\}_{i=1}^N$ represent received signal envelopes in
an isotropic scattering radio channel exposed to the Doppler
Effect, they must be considered as time-correlated random
processes with some resulting Doppler spectrum. This Doppler
spectrum differs depending on whether f\-ixed-to-mobile channel
\cite{11}-\cite{12} or mobile-to-mobile channel
\cite{13}-\cite{14} appears in the wireless
communications system. In both cases, it was found that time
derivative of $i$-th envelope   is independent from the envelope
itself, and follows the Gaussian PDF \cite{11}-\cite{14}
\begin{equation}\label{2}
f_{\dot X_i}(\dot x )=\frac{1}{\sqrt{2\pi}\sigma_{\dot X_i }}
\exp\Big(-\frac{\dot x^2 }{2\sigma_{\dot X_i }^2}\Big) ,
\end{equation}
with variance calculated as
\begin{equation}\label{3}
\sigma_{\dot X_i }^2=\pi^2\Omega_i f_i^2 \,.
\end{equation}

If envelope $X_i$ is formed on a f\-ixed-to-mobile channel, then
$f_i=f_{mi}$ where $f_{mi}$ is the maximum Doppler frequency shift
induced by the motion of the mobile station \cite{11}-\cite{12}.
If envelope $X_i$ is formed on a mobile-to-mobile channel, then
\begin{equation}\label{3a}
f_i=\sqrt{f_{mi}^{'2}+f_{mi}^{''2}} \,.
\vspace{-1.0mm}
\end{equation}
where $f_{mi}^{'}$ and $f_{mi}^{''}$ are the maximum Doppler
frequency shifts induced by the motion of both mobile stations
(i.e., the transmitting and the receiving stations, respectively)
\cite{14}. It is important to underline that the maximum Doppler
frequency in a f\-ixed-to-mobile channel is $f_{d \max}=f_{mi}$,
whereas the maximum Doppler frequency in a mobile-to-mobile
channel is $f_{d \max}=f_{mi}^{'}+f_{mi}^{''}$. The above results
are essential in deriving the second-order statistical parameters
of individual envelopes, as the LCR and the AFD \cite{11},
\cite{12}, \cite{14}.

Below, we derive exact and approximate solutions for both of the
above parameters for product of $N$ Rayleigh envelopes,
\begin{equation}\label{4}
Y(t)=\prod_{i=1}^N X_i(t) \,.
\vspace{-1.0mm}
\end{equation}
We denote $Y(t)$ as $N*$Rayleigh random process or, at any given
moment $t$, $N*$Rayleigh random variable, following the
def\-inition given in \cite{15}.

For some specif\-ied value $\{X_i\}_{i=1}^N = \{x_i\}_{i=1}^N$,
the product $Y$ is f\-ixed to the specif\-ic value
$y=\prod_{i=1}^N x_i$. The LCR of $Y$ at threshold $y$ is
def\-ined as the rate at which the random process crosses level
$y$ in the negative direction \cite{11}. To extract LCR, we need
to determine the joint probability density function (PDF) between
$Y$ and $\dot Y$, $f_{Y\dot Y}(y,\dot y)$, and to apply the Rice's
formula [11, Eq. (2.106)],
\begin{equation}\label{6}
N_Y(y)=\int_0^\infty \dot y f_{Y\dot Y} (y,\dot y) d\dot y \,.
\vspace{-0.0mm}
\end{equation}
Our method does not require explicit determination of $f_{Y\dot
Y}(y,\dot y)$ in order to determine analytically the LCR of the
$N*$Rayleigh random process, as presented below.

F\-irst, we need to f\-ind the time derivative of (\ref{4}), which
is
\begin{equation}\label{7}
\dot Y=Y\sum_{i=1}^N\frac{\dot X_i}{X_i} \,.
\vspace{-0.0mm}
\end{equation}

Conditioning on the f\-irst $N-1$ envelopes $\{X_i\}_{i=1}^{N-1} =
\{x_i\}_{i=1}^{N-1}$, we have the conditional joint PDF $Y$ and
$\dot Y$ written as $f_{Y\dot Y|X_1\cdot\cdot\cdot X_{N-1}}(y,\dot
y|x_1,...,x_{N-1})$. This conditional joint PDF can be averaged
with respect to the joint PDF of the $N-1$ envelopes
$\{X_i\}_{i=1}^{N-1}$ to produce the required joint PDF,
\begin{eqnarray}\label{8}
f_{Y\dot Y}(y,\dot y) \qquad \qquad \qquad \qquad \qquad \qquad \qquad \qquad \qquad \,\, \nonumber \\
 =  \int_{x_1=0}^\infty \cdots\int_{x_{N-1}=0}^\infty f_{Y\dot Y | X_1\cdots X_{N-1}}(y,\dot y | x_1,...,
 x_{N-1}) \nonumber \\
\times \, f_{X_1}(x_1)\cdots f_{X_{N-1}}( x_{N-1})dx_1\cdots dx_{N-1}
\vspace{-0.0mm}
\end{eqnarray}
where to derive (\ref{8}) the mutual independence of the $N-1$
envelopes is used.

The conditional joint PDF $f_{Y\dot Y|X_1\cdot\cdot\cdot
X_{N-1}}(y,\dot y|x_1,...,$ $x_{N-1})$ can be further simplif\-ied
by setting $Y = y$ and using the total probability theorem,
\begin{eqnarray}\label{9}
f_{Y\dot Y | X_1\cdots X_{N-1}}(y,\dot y | x_1,...,
x_{N-1}) \qquad \qquad \qquad \qquad \nonumber \\
=f_{\dot Y | Y X_1\cdots X_{N-1}}(\dot y |y, x_1,...,
x_{N-1}) \qquad \qquad \quad \nonumber \\
\times \, f_{Y| X_1\cdots X_{N-1}}(y | x_1,..., x_{N-1}) \,\,,
\end{eqnarray}
where each of the two multipliers in (\ref{9}) can be determined
from the above def\-ined individual PDFs and their parameters.

Based on (\ref{7}), the conditional PDF $f_{\dot Y|Y
X_1\cdot\cdot\cdot X_{N-1}}(\dot y|y,x_1,$ $...,x_{N-1})$ can be
easily established to follow the Gaussian PDF with zero mean and
variance
\begin{eqnarray}\label{10}
\sigma_{\dot Y | Y X_1\cdots X_{N-1}}^2=\left(
y^2\sum_{i=1}^{N-1}\frac{\sigma_{\dot X_i}^2}{x_i^2}+ \sigma_{\dot
X_N}^2\prod_{i=1}^{N-1}x_i^2\right) \qquad \quad \nonumber \\
=\sigma_{\dot X_N}^2\left[ 1+y^2\left(\prod_{i=1}^{N-1}
\frac{1}{x_i^2}\right) \sum_{i=1}^{N-1}\frac{\sigma_{\dot
X_i}^2}{\sigma_{\dot X_N}^2}
\frac{1}{x_i^2}\right]\prod_{i=1}^{N-1}x_i^2 \,.
\end{eqnarray}

The conditional PDF of $Y$ given $\{X_i\}_{i=1}^{N-1} =
\{x_i\}_{i=1}^{N-1}$ that appears in (\ref{9}) is easily
determined in terms of the PDF of the remaining $N$-th envelope,
\begin{eqnarray}\label{11}
f_{Y|X_1\cdots X_{N-1}}(y|x_1,...,
x_{N-1}) \qquad \qquad \qquad \qquad \qquad \nonumber \\
=f_{X_N}\left(y \prod_{i=1}^{N-1} \frac{1}{x_i}\right)
\prod_{i=1}^{N-1}\frac{1}{x_i} \qquad
\end{eqnarray}
Introducing (\ref{9}) and (\ref{11}) into (\ref{8}), then
(\ref{8}) into (\ref{6}) and changing the orders of the
integration, we obtain
\begin{eqnarray}\label{12}
N_Y(y) = \int_{x_1=0}^\infty \cdots\int_{x_{N-1}=0}^\infty
\qquad \qquad \qquad \qquad \qquad \qquad \nonumber \\
\left(\int_{\dot y=0}^\infty \dot y f_{\dot Y | Y X_1\cdots
X_{N-1}}(\dot y |y, x_1,..., x_{N-1})d\dot y \right)
\prod_{i=1}^{N-1}\frac{1}{x_i} \qquad \nonumber\\
f_{X_N}\left (y\prod_{i=1}^{N-1}\frac{1}{x_i}\right )
f_{X_1}(x_1)\cdots f_{X_{N-1}}(x_{N-1})dx_1\cdots dx_{N-1}
\end{eqnarray}
The bracketed integral in (\ref{12}) is found using (\ref{10}) as
\begin{eqnarray}\label{13}
\int_{0}^\infty \dot y f_{\dot Y | Y X_1\cdots X_{N-1}}(\dot y |y,
x_1, \cdots , x_{N-1})d\dot y
=\frac{\sigma_{\dot Y | Y X_1\cdots
X_{N-1}}} {\sqrt{2\pi}}
\end{eqnarray}
By substituting (\ref{1}) and (\ref{13}) into (\ref{12}), we
obtain the exact formula for the LCR as
\begin{eqnarray}\label{14}
N_Y(y)= \frac{\sigma_{\dot X_N}}{\sqrt{2\pi}}\frac{2^Ny}{\Phi} \qquad \qquad \qquad \qquad \qquad \qquad \qquad \qquad \nonumber \\
\times \int_{x_1=0}^{\infty} \cdots\int_{x_{N-1}=0}^\infty\left[
1+y^2\left(\prod_{i=1}^{N-1}
\frac{1}{x_i^2}\right)\sum_{i=1}^{N-1}\frac{\sigma_{\dot
X_i}^2}{\sigma_{\dot X_N}^2} \frac{1}{x_i^2}\right]^{1/2}
\nonumber \\
\times\exp\left[-\left(\frac{y^2}{\Omega_N}\prod_{i=1}^{N-1}
\frac{1}{x_i^2}+\sum_{i=1}^{N-1}\frac{x_i^2}{\Omega_i}\right)\right]dx_1\cdots
dx_{N-1}, \qquad
\end{eqnarray}
where
\begin{equation}\label{15}
\Phi =\prod_{k=1}^{N}\Omega_k
\end{equation}

In principle, (\ref{14}) together with (\ref{15}) provide an exact
analytical expression for the LCR of the product of the product of
$N$ Rayleigh envelopes (i.e., $N*$Rayleigh random process
\cite{15}). However, (\ref{14}) becomes computationally attractive
only for small values of $N$, where it is possible to apply a
numerical computation method (as Gaussian-Hermite quadrature).

Note that, (\ref{14}) is transformed into a single integral when
$N=2$, which, after introducing (\ref{3}) for $i = 1, 2$ and
changing integration variable $x$ with new variable $t$ according
$x=y/t$, reduces to the known result [9, Eq. (17)].

The AFD of $Y$ at threshold $y$ is def\-ined as the average time
that the $N*$Rayleigh random process remains below level $y$ after
crossing that level in the downward direction,
\begin{equation}\label{16}
T_Y(y)=\frac{F_Y(y)}{N_Y(y)} ,
\end{equation}
where $F_Y(\cdot)$ denotes the cumulative distribution function
(CDF) of $Y$. Fortunately, $F_Y(\cdot)$ was derived recently in
closed-form [14, Eq. (7)], as
\begin{equation}\label{17}
F_Y(y)=G_{1,N+1}^{N,1} \left[\frac{y^2}{\Phi}\Bigg |
\begin{array}{cc}
 \qquad 1 \\
 \underbrace{1,1,\cdots ,1}_N, 0
\end{array}
 \right],
\end{equation}
where $G[\cdot]$ is the Meijer's $G$-function [15, Eq. (9.301)].

\subsection{An Approximate Solution for the LCR}
Next, we present a tight closed-form approximation of (\ref{14})
using the multivariate Laplace approximation theorem [16, Chapter
IX.5], \cite{18} for the Laplace-type integral
\begin{equation}\label{18}
J(\lambda)=\int_{\textbf{x} \in D} u(\textbf{x}) \exp(-\lambda
h(\textbf{x}))d\textbf{x} ,
\end{equation}
where $u$ and $h$ are real-valued multivariate functions of
$\mathbf{x}=[x_1, \cdots , x_{N-1}]$, $\lambda$ is a real
parameter and $D$ is unbounded domain in the multidimensional
space $R ^{N-1}$.

A comparison of (\ref{14}) and (\ref{18}) yields
\begin{equation}\label{19}
u(\textbf{x})=\left[1+y^2\left(\prod_{i=1}^{N-1}\frac{1}{x_i^2}\right)\sum_{i=1}^{N-1}\frac{\sigma_{\dot
X_i}^2}{\sigma_{\dot X_N }^2}\frac{1}{x_i^2}\right]^{1/2} ,
\end{equation}

\begin{equation}\label{20}
h(\textbf{x})=\frac{y^2}{\Omega_N}\prod_{i=1}^{N-1}\frac{1}{x_i^2}+\sum_{i=1}^{N-1}\frac{x_i^2}{\Omega_i},
\end{equation}
and $\lambda=1$. Note, that in the case of (\ref{14}), all the
applicability conditions of the theorem are fulf\-illed. Namely,
within the domain of interest $D$, the function $h(\mathbf{x})$
has a single interior critical point $\tilde \textbf{x}=[\tilde
x_1,\cdots ,\tilde x_{N-1}]$, where
\begin{equation}\label{21}
\tilde{x_i}=y^{1/N}\frac{\Omega_i^{1/2}}{\Phi^{1/(2N)}} ,\qquad
1\leq i\leq N-1 ,
\end{equation}
which is obtained from solving the set of equations $\partial h
/\partial x_i =0$, where $1\leq i\leq N-1$. The Hessian
$(N-1)\times (N-1)$ square matrix $\mathbf A$, def\-ined by [15,
Eq. (14.314) ], is written as
\begin{equation}\label{22}
\mathbf{A}=\left[
\begin{array}{cccc}
8/\Omega_1 & 4/\sqrt{\Omega_1\Omega_2} & \cdots & 4/\sqrt{\Omega_1\Omega_{N-1}}\\
4/\sqrt{\Omega_2\Omega_1} & 8/\Omega_2 & \cdots & 4/\sqrt{\Omega_2\Omega_{N-1}}\\
. & . & \cdots & .\\
4/\sqrt{\Omega_{N-1}\Omega_1} & 4/\sqrt{\Omega_{L-1}\Omega_2} &
\cdots & 8/\Omega_{N-1}
\end{array}
\right ]
\end{equation}
By using induction, it is easy to determine that the $N-1$
eigenvalues of $\mathbf A$ are calculated as $\mu_i=4/\Omega_i$
for $1\leq i \leq N-2$, and $\mu_{N-1}=4N/\Omega_{N-1}$. Thus, all
eigenvalues of $\mathbf A$ are positive, which, by def\-inition,
means that the matrix $\mathbf A$ is positive def\-inite. By means
of the second derivative test, since the Hessian matrix $\mathbf
A$ is positive def\-inite at point $\tilde \mathbf x$, $h(\mathbf
x)$ attains a local minimum at this point (which in this case is
the absolute minimum in the entire domain $D$).

At this interior critical point $\tilde \mathbf x$,
\begin{equation}\label{23}
u(\mathbf{\tilde{x}})=\left (1+\sum_{i=1}^{N-1}\frac{\sigma_{\dot
X_i}^2}{\sigma_{\dot
X_N}^2}\frac{\Omega_N}{\Omega_i}\right)^{1/2}= \left (
1+\sum_{i=1}^{N-1}\frac{f_i^2}{f_N^2}\right)^{1/2} ,
\end{equation}
\begin{equation}\label{24}
h(\mathbf{\tilde{x}})=N\left (\frac{y^2}{\Phi}\right)^{1/N} \, ,
\end{equation}
where (\ref{23}) is obtained using (\ref{3}). Now,
 it is possible to approximate (\ref{18}) for large
$\lambda$ as
\begin{eqnarray}\label{25}
J(\lambda)\approx \left(
\frac{2\pi}{\lambda}\right)^{(N-1)/2}\left[ \frac{1}{\det
(\mathbf{A})}\left(1+\sum_{i=1}^{N-1}\frac{f_i^2}{f_N^2}\right
)\right ]^{1/2} \nonumber \\
\times \exp\left (-\lambda N\frac{y^{2/N}}{\Phi^{1/N}}\right) .
\end{eqnarray}
It is well-know that the determinant of the square matrix is equal
to the product of its eigenvalues. Hence, $\mathbf A$ can be
written as
\begin{equation}\label{26}
\det(\mathbf A)=\frac{N2^{2(N-1)}}{\prod_{k=1}^{N-1}\Omega_k}
=\frac{\Omega_N N2^{2(N-1)}}{\Phi} .
\end{equation}
Although approximation (\ref{25}) is proven for large $\lambda$
\cite{17}-\cite{18}, it is often applied when $\lambda$ is small
and is observed to be very accurate as well. Similarly to
\cite{19}, we apply the theorem for  $\lambda = 1$. Therefore, the
approximate closed-form solution for the LCR of $N*$Rayleigh
random process $Y$ at threshold $y$ is
\begin{eqnarray*}\label{27}
N_Y(y) \approx \frac{\sigma_{\dot
X_{N}}}{\sqrt{2\pi}}\frac{2^Ny}{\Phi}J(1) = \frac{2y
(2\pi)^{N/2-1} \sigma_{\dot X_N}}{\Omega_N^{1/2} \Phi^{1/2}}
\qquad \qquad \qquad \nonumber \\
\times \left[\frac{1}{N} \left(
1+\sum_{i=1}^{N-1}\frac{f_i^2}{f_N^2}\right)\right]^{1/2}
 \exp\left(-N\frac{y^{2/N}}{\Phi^{1/N}}\right) \qquad \nonumber\\
\end{eqnarray*}
\begin{equation}\label{27}
= \left(\frac{1}{N}\sum_{i=1}^N
f_i^2\right)^{1/2}\frac{(2\pi)^{N/2}y}{\Phi^{1/2}}
\exp\left(-N\frac{y^{2/N}}{\Phi^{1/N}}\right) \,. \quad
\end{equation}

The numerical results presented in Section IV validate the high
accuracy of the Laplace approximation applied for our particular
case.

Combining (\ref{17}) and (\ref{27}) into (\ref{16}), the AFD of
the $N*$Rayleigh random process $Y$ at threshold $y$ is
approximated as
\begin{eqnarray}\label{27a}
T_Y(y)\approx\left(\frac{1}{N}\sum_{i=1}^{N}f_i^2\right)^{-1/2}\frac{\Phi^{1/2}}{(2\pi)^{N/2}}\frac{1}{y}
\qquad \qquad \qquad \qquad \nonumber \\
\times G_{1,N+1}^{N,1} \left[\frac{y^2}{\Phi}\Bigg |
\begin{array}{cc}
& \qquad 1 \\
& \underbrace{1,1,\cdots ,1}_N, 0
\end{array}
 \right]\exp\left(N\frac{y^{2/N}}{\Phi^{1/N}}\right).
\end{eqnarray}

\section{Second Order Statiscs of Multihop Transmission}
Next, we apply the important theoretical result of the previous
Section to analyze the second order statistics of the multihop
relay fading channel.
\vspace{-1.5mm}
\subsection{System Model}

We now consider a multihop wireless communications system,
operating over i.n.i.d f\-lat fading channels. Source station $S$
communicates with destination station $D$ through $N-1$ relays
$T_1$, $T_2$,..., $T_{N-1}$, which act as intermediate stations
from one hop to the next. These intermediate stations are employed
with non-regenerative relays with f\-ixed gain $G_i$ given by
\begin{equation}\label{30}
G_i^2=\frac{1}{C_iW_{0,i}}
\end{equation}
with $G_0=1$ and $C_0=1$ for the source $S$. In (\ref{30}),
$W_{0,i}$ is the variance of the Additive White Gaussian Noise
(AWGN) at the output of the $i$-th relay, and $C_i$ is a constant
for the f\-ixed gain $G_i$.

Assume that terminal $S$ is transmitting a signal $s(t)$ with an
average power normalized to unity. Then, the received signal at
the f\-irst relay, $T_1$, at moment $t$, can be written as
\begin{equation}\label{31}
r_1(t)=\alpha_1(t)s(t)+w_1(t) \,,
\end{equation}
where $\alpha_1(t)$ is the fading amplitude between $S$ and $T_1$,
and $w_1(t)$ is the AWGN at the input of $T_1$ with variance
$W_{0,1}$. The signal $r_1$ is then multiplied by the gain $G_1$
of the relay $T_1$ and re-transmitted to relay $T_2$. Generally,
the received signal at the $k$-th relay $T_k$ ($k=1, 2,..., N-1$)
is given by
\begin{equation}\label{33}
r_k(t)=G_{k-1}\alpha_k(t)r_{k-1}(t)+w_k(t) \,,
\end{equation}
resulting in a total fading amplitude at the destination node $D$,
given by
\begin{equation}\label{34}
\alpha(t)=\prod_{i=1}^N\alpha_i(t)G_{i-1} \,.
\end{equation}

\subsection{LCR and AFD of Multihop Transmissions}

If the fading amplitude received at node $T_i$, $\alpha_i(t)$, is
a time-correlated (due to mobility of $T_{i-1}$ and/or $T_i$)
Rayleigh random process, distributed according to (\ref{1}) with
mean power $\hat \Omega_i=E\{\alpha_i^2(t)\}$, then the $i$-th
element of the product in (\ref{34}), $X_i(t)=\alpha_i(t)G_{i-1}$,
is again a time-correlated Rayleigh random process, distributed
according to (\ref{1}) with mean power $\Omega_i=\hat \Omega_i\,
G_{i-1}^2$.

Comparing (\ref{4}) and (\ref{34}), we realize that the total
fading amplitude at the destination station $D$ (i.e., the
received desired signal without the AWGN) is described as the
$N*$Rayleigh random process $Y(t)=\alpha(t)$, whose average LCR
and AFD are determined in the previous Section.

If all stations are assumed mobile with maximum Doppler frequency
shifts $f_{mS}, f_{mD}$, $f_{mi} (1\leq i\leq N-1)$ for the
source $S$, destination $D$ and relays, respectively, then for the
$i$-th hop $f_i^2=f_{m(i-1)}^2+f_{mi}^2$ with $f_{m0}=f_{mS}$ and
$f_{mN}=f_{mD}$, and
\begin{equation}\label{35}
\sum_{i=1}^N f_i^2=f_{mS}^2+2\sum_{i=1}^{N-1} f_{mi}^2+f_{mD}^2
\,.
\end{equation}
\noindent Combining (\ref{27}) and (\ref{35}), we obtain
approximate solution for the average LCR of the total fading
amplitude $\alpha$ at the output of a multihop non-regenerative
relay transmission system, \vspace{-1.5mm}
\begin{eqnarray}\label{36}
N_{\alpha}(\alpha)\approx
\left[\frac{1}{N}\left(f_{mS}^2+2\sum_{i=1}^{N-1}
f_{mi}^2+f_{mD}^2\right)\right]^{1/2} \qquad \quad \nonumber \\
\times \frac{(2\pi)^{N/2}\alpha}{\Phi^{1/2}}
\exp\left(-N\frac{\alpha^{2/N}}{\Phi^{1/N}}\right) \,,
\end{eqnarray}
where $\Phi$ is given by (\ref{15}). We see that (\ref{36})
approximates the average LCR of the total fading amplitude for
arbitrary mean power of the fading amplitudes $\hat \Omega_i$,
arbitrary relay gains $G_i$ and arbitrary maximal Doppler shifts
$f_{mi}$.

Note that, for $N=2$, (\ref{36}) is an eff\-icient closed-form
alternative to the corresponding one [9, Eq. (17)] for the
dual-hop case, which is shown in next section to be highly
accurate.

\section{Numerical Results and Discussion}
In this section, we provide some illustrative examples for the
average LCR and AFD of the fading gain process of the received
desired signal at the destination of the multihop non-regenerative
relay transmission system model from Section III. The numeric examples obtained from the
derived approximate solutions are validated by extensive
Monte-Carlo simulations.

We considered a multihop system consisted of a source
terminal $S$, 4 relays, and a destination terminal $D$. The
f\-ixed-gain relays are assumed semi-blind with gains in Rayleigh
fading channel calculated as [2, Eq. (15)] and [6, Eq.
(19)]
\begin{equation}\label{38a}
G_{i,sb}^2=\frac{1}{\hat
\Omega_i}\exp\left(\frac{1}{\bar\gamma_i}\right)\Gamma\left(0,\frac{1}{\bar\gamma_i}\right),
\end{equation}
where $\bar \gamma_i=\hat \Omega_i/W_{0,j}$ is the mean SNR on the
i-th hop, and $\Gamma(\cdot,\cdot)$ is the incomplete Gamma
function. Relay gain calculated according to (\ref{38a}) assures
mean power consumption equal to that of a CSI-assisted relay,
whose gain inverts the fading effect of the previous hop while
limiting the output power at moments with deep fading.

Depending on the stations' mobility, we used two different 2D
isotropic scattering models for the Rayleigh radio channel on each
hop of the multihop transmission system. For the f\-ixed-to-mobile
channel (hop), we used the classic Jakes channel model
\cite{11}-\cite{12}. For the mobile-to-mobile channel (hop), we
used the Akki and Habber's channel model \cite{13}-\cite{14}. The
Monte-Carlo simulations of the latter were realized by using the
sum-of-sinusoids method proposed in \cite{20}-\cite{21}.

More precisely, all mobile stations are assumed to induce same
maximal Doppler shifts $f_m$, while the destination $D$ is
f\-ixed. For all hops, $\hat \Omega_i=\hat \Omega$ and
$W_{0,i}=W_0$. Thus, $\bar\gamma_{i}=\bar\gamma$,
$G_{i,sb}=G_{sb}$, and the mean of Rayleigh random process
$X_i(t)=\alpha_i(t)G_{i-1,sb}$ is calculated as
\begin{equation}\label{38b}
\Omega_i =
\exp\left(\frac{1}{\bar\gamma}\right)\Gamma\left(0,\frac{1}{\bar\gamma}\right)=
\Omega \,\,, 2\leq i \leq N
\end{equation}
whereas $\Omega_1=\hat \Omega$ is selected independently from the
AWGN, since $G_0=1$. In this case,
\begin{equation}\label{38c}
\Phi=\hat\Omega\,\exp\left(\frac{N-1}{\bar\gamma}\right)\left[\Gamma\left(0,\frac{1}{\bar\gamma}\right)\right]^{N-1}
\end{equation}
Note that, when introducing above scenario into (\ref{36}),
$\alpha$ and $\hat \Omega$ appear together as $\alpha/\sqrt{\hat
\Omega}$.

\begin{figure}
\centering
\includegraphics[width=3.2in]{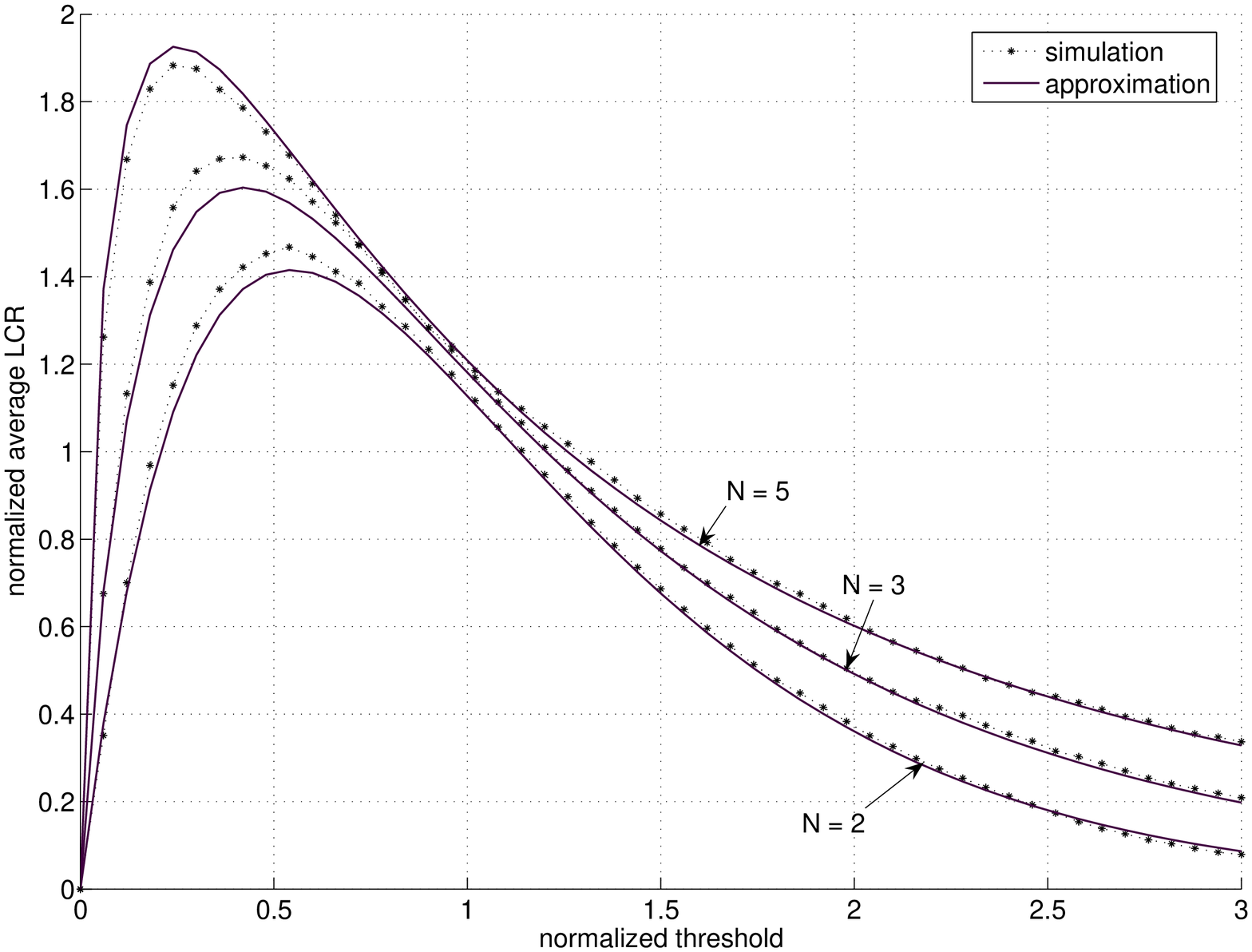}
\vspace{-5mm} \caption {Average LCR, $\hat \gamma_i=\hat \gamma =
5$ dB} \label{fig_1} \vspace{-4mm}
\end{figure}

\begin{figure}
\centering
\includegraphics[width=3.2in]{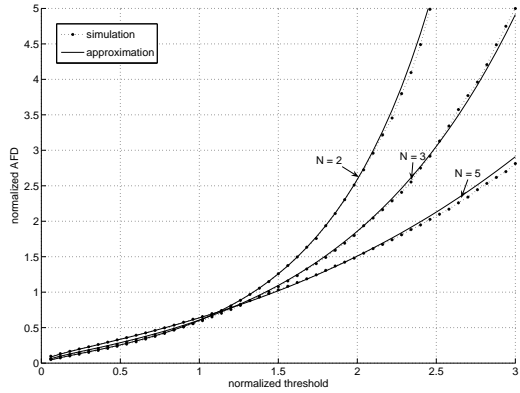}
\vspace{-5mm} \caption {AFD, $\hat \gamma_i=\hat \gamma = 5$ dB}
\label{fig_2} \vspace{-4mm}
\end{figure}
Figs. 1-4 depict the received signal's normalized LCR
($N_{\alpha}/f_m$) or normalized AFD ($T_{\alpha}f_m$) versus the
normalized threshold ($\alpha/\sqrt{\hat \Omega}$) at 3 different
stations along the multihop transmission system: at relay $T_2$
(curve denoted by $N = 2$), at relay $T_3$ (curve denoted by $N =
3$) and at the destination $D$ (curve denoted by $N = 5$). All
comparative curves show an excellent match between the approximate
solution and the Monte-Carlo simulations.


\begin{figure}
\centering
\includegraphics[width=3.2in]{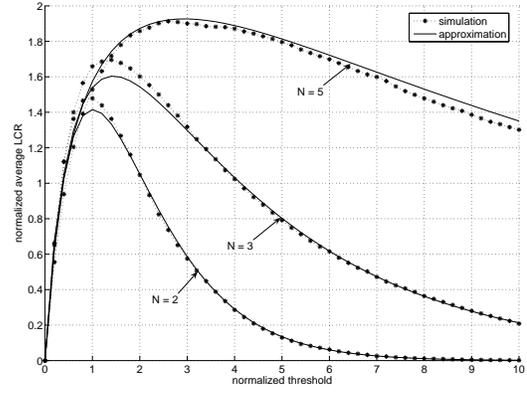}
\vspace{-5mm} \caption{Average LCR, $\hat \gamma_i=\hat \gamma =
20$ dB} \label{fig_3} \vspace{-3mm}
\end{figure}

\begin{figure}
\centering
\includegraphics[width=3.2in]{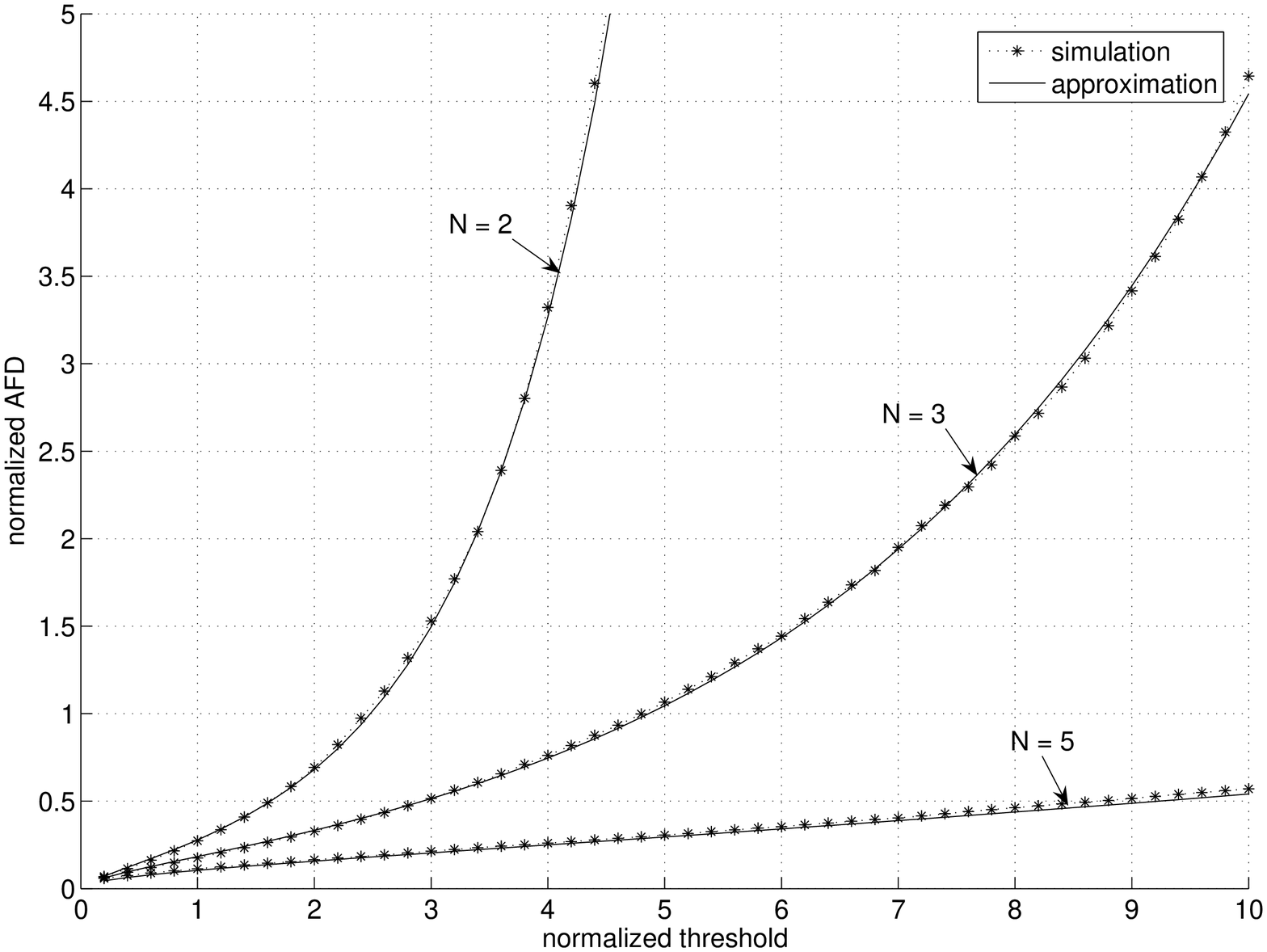}
\vspace{-5mm} \caption {AFD, $\hat \gamma_i=\hat \gamma = 20$ dB}
\label{fig_4} \vspace{-6mm}
\end{figure}


\end{document}